\magnification=1200
\baselineskip=20 pt

\def\lv{\lambda_v}
\def\lh{\lambda_h}

\def\modphi{\vert \phi \vert}
\def\rc{r_c}
\def\vv{v_v}
\def\vh{v_h}
\def\ep{\epsilon}
\def\del{\partial}
\def\tphi{\tilde{\phi}}

\centerline{\bf Radion stabilization in the Randall-Sundrum model}

\centerline{\bf with quadratic and quartic potentials}

\vskip 1 true in

\centerline{\bf Uma Mahanta} 

\centerline{\bf Mehta Research Institute}

\centerline{\bf Chhatnag Road, Jhusi}

\centerline{\bf Allahabad-211019, India}

\vskip .5 true in

\centerline{\bf Abstract}

In this report we investigate the Goldberger-Wise (GW) mechanism of
radion stabilization with quartic potential on the hidden brane
and quadratic potential on the visible brane. The advantage of our
simplified scenario over the original GW mechanism is that the modulus
potential can be evaluated for finite $\lv$ and $\lh$. This enables us 
to probe how the modulus potential behaves over the entire range of
$\lh$. By staying away from the GW limit of ${k \over \lv v_v^2}
\rightarrow $0
and  ${k \over \lh v_h^2}\rightarrow $0 
we show that it is possible to choose the 
parameters of the model so that the potential exhibits a
minimum at $k\rc \approx$ 12 and this adjustment
does not involve any extreme fine tuning of parameters.

\vfill\eject

Several radical proposals based on higher dimensional theories have been
recently put forward to explain the large hierarchy between the weak
scale and the Planck scale. Among them the Randall-Sundrum (RS) [1]
scenario
is particularly attractive since it explains the hierarchy in trems of
a small extra dimensions. Unlike theories with large extra dimensions,
in the RS model there is no large hierarchy between the compactification
scale ${1\over r_c}$ and the fundamental Planck mass M. The reason behind 
this difference is that the hierarchy is explained in terms of an 
exponential warp factor that appears in the non-factorizable metric
of the five dimensional RS world.

$$ds^2= e^{-2kr_c\modphi }\eta_{\mu\nu} dx^{\mu}dx^{\nu}-r_c^2 d\phi^2
.\eqno(1)$$

Here k is a parameter of the order of M. $-\pi\le \phi \le $ is the 
coordinate of the single ${S_1\over Z_2}$ orbifold extra dimension.
The points $(x, \phi )$ and  $(x, -\phi )$ are therefore identified.
The hidden 3 brane is located at $\phi =0$ and the visible brane
is located at $\phi =\pi$. 

In the RS scenario the compactification radius $r_c$ was associated 
with the vacuum expectation value (vev) of a four dimensional massless
scalar field T(x).
However T(x) had zero potential and $r_c$ was not stabilized by some 
dynamics. Goldberger and Wise (GW) [2] showed that it is possible to
generate a potential $V( \rc )$ for the modulus field by introducing a 
bulk scalar field $\chi (x, \phi )$ with interaction potentials localized
on the two branes. They also showed that the minimum of the modulus
potential can be adjusted to occur at $k\rc \approx $ 12 without fine tuning
the parameters of the model. In their original proposal GW assumed the
interaction potentials on the two branes to be quartic functions of $\chi$.
As a result they could determine the modulus potential $V(\rc )$ only in 
the limit of infinite $\lv$ and $\lh$. For large but finite $\lv$
and $\lh$ the shifts in the vevs $\delta\chi (\pi )$ and $\delta\chi (0)$
from their values $v_v$ and $v_h$ at infinite $\lv$ and $\lh$ are given by
[2]

$$\delta\chi (\pi )=-{k\over \lv v_v^2} (\vv -\vh e^{-\ep k\rc\pi}).
\eqno(2a)$$

$$\delta\chi (0)=-{k\over \lh \vh^2} e^{-(4+\ep )k\rc\pi }
(\vv -\vh e^{-\ep k\rc\pi}).\eqno(2b)$$

Therefore the GW limit assumes that ${k\over \lv v_v^2}$ and
${k\over \lh \vh^2}$ have been tuned to very small values which 
may not stable under fluctuations in the background
metric or quantum corrections.
 It is therefore worthwhile
to avoid this fine tuning and determine the modulus potential for
finite $\lv$ and $\lh$. However for quartic brane potentials with
finite $\lv$ and $\lh$ it is difficult to determine the background
field configuartion $\chi (\phi )$ that satisfies the appropriate
boundary conditions on the branes analytically. Therefore in this 
report we shall consider a simplified scenario with quartic interaction 
potential for $\chi $ on the hidden brane and quadratic potential
on the visible brane. This set up will enable us to determine the
modulus potential analytically for finite $\lv$ and $\lh$. We find that 
the modulus potential $V(\rc )$ for this scenario can be adjusted
to yield a minimum both for finite and infinite $\lh$ at 
$k\rc \approx $ 12 without fine tuning the parameters of the model.

The action for our scenario is given by

$$S_b={1\over 2}\int d^4x\int^{-\pi}_{\pi}d\phi
\sqrt {G}(G^{AB}\partial_A\chi\partial_B\chi -m^2\chi^2).\eqno(3a)$$

$$S_h=-\int d^4x \int d\phi \sqrt{-g_h}\lh (\chi^2-v_h^2)^2
{\delta (\phi)\over \rc }.\eqno(3b)$$

$$S_v=-\int d^4x \int d\phi \sqrt{-g_v}\lv v_v^2 (\chi^2-v_v^2 )
{\delta (\phi -\pi ) \over \rc }.\eqno(3c)$$

In this work we shall ignore the variation of $\chi$ parallel to the 
3 branes. This is justified since we are interested only in the
stability of the size of the extra dimension or
the interbrane separation.
It can be shown that away from the boundaries $\phi = 0, \pi $
the  classical configuration of the bulk field $\chi$ is given 
by

$$\chi ( \phi )=Ae^{(2+\nu )\sigma}+Be^{(2-\nu )\sigma}.\eqno(4)$$

where $\sigma =k\rc\vert \phi\vert $ and $\nu =\sqrt{4+{m^2\over k^2}}$.
In particular we shall consider the case $m\ll k$ so that
$\nu\approx 2+\ep$ where $\ep\approx {m^2\over 4k^2}$.
The constants A and B are determined by the boundary conditions imposed
by the interaction potentials on the two branes. Using the relations
$\chi^{\prime}(0+\ep )=-\chi^{\prime}(0-\ep )$ and 
$\chi^{\prime}(\pi +\ep )=-\chi^{\prime}(\pi -\ep )$
the boundary conditions can be written as 

$$k[(2+\nu )e^{(2+\nu )k\rc\pi}A+(2-\nu )e^{(2-\nu )k\rc\pi}B]
+\lv v_v^2\chi (\pi ) =0.\eqno(5)$$

and

$$k[(2+\nu )A + (2-\nu )B]-2\lh \chi (0)[\chi^2 (0)-v_h^2]=0.\eqno(6)$$

Here $\chi (0)=A+B$ and $\chi (\pi)=Ae^{(2+\nu )k\rc\pi}
+Be^{(2-\nu )k\rc\pi}$. In order to solve (5) and (6) for A and
B we shall assume that $k\rc >1$ so that we can neglect higher powers
of $e^{-2\nu k\rc\pi}$. We then get

$$A=\pm \sqrt {a} b e^{-2\nu k\rc\pi} [1+ \{ c(1+{2+\nu\over\ep})-1\}
be^{-2\nu k\rc\pi} ].\eqno(7)$$

and

$$B=\pm \sqrt{a} [1+ \{ c(1+{2+\nu\over\ep})-1\}
b e^{-2\nu k\rc\pi}].\eqno(8)$$ 

where $a=v_h^2-{k\ep\over 2\lh}$, $b={k\ep-\lv v_v^2\over k(2+\nu )+
\lv v_v^2}$ and $c={k\ep\over 4\lh}{1\over v_h^2-{k\ep\over 2\lh}}$.
Putting the solution (4) back into the action and integrating over
$\phi$ yields the following four dimensional potential for the modulus

$$V(\rc )=k\ep (v_h^2-{k\ep\over 4\lh})+ka[ (2+\nu )b^2
-\ep (1+2b)]\tphi^{4+2\ep}+\lv v_v^2[(1+b)^2 \tphi^{4+2\ep}
-v_v^2\tphi^4].\eqno(9)$$

where $\tphi =e^{-k\rc\pi}$
In the above expression for $V(\rc )$ we have 
retained only terms up to order $e^{-2\nu k\rc \pi}$. Since 
 $e^{-2\nu k\rc \pi}\ll 1$ even for $k\rc \approx $ 1 we can trust
the above expression for $V(\rc )$ as long as $k\rc\ge $1. 
Note that as $\lv\rightarrow \infty $, $b\approx -(1-{(2+\nu )k\over 
\lv v_v^2})$ and 
 $\chi (\pi)\rightarrow 0$.
The visible brane potential therefore approaches the value $-\lv
v_v^4 $. Since the visible brane potential becomes unbounded from
 below for $\lv\rightarrow \infty$ we shall restrict ourselves to finite
values of $\lv$ only.

The above expression for $V(\rc )$ will exhibit a non-trivial minimum
provided the coefficient of the $\tphi^{4+2\ep}$ term is positive. 
In this case as $\tphi$ increases from zero the $\tphi^4$ term first 
dominates causing $V(\rc )$ to decrease from its value 
$k\ep (v_h^2-{k\ep\over 4\lh})$ at $\tphi =0$. However with gradually
increasing $\tphi$, the $\tphi^{4+2\ep}$ term ultimately
begins to dominate causing $V(\rc )$ to increase and exhibit a 
non-trivial minimum. In the following we shall assume that $\lv$
and $\lh$ are finite. In particular we shall consider the case
where ${k \ep over \lv v_v^2}$ and  ${k\ep\over \lh v_h^2}$
are of order unity. In other words we shall study the minima
of $V(\rc )$ by staying away from the GW limit of ${k \over \lv v_v^2}
\rightarrow $0 and  ${k \over \lh v_h^2}\rightarrow $0.

Let us choose the parameters so that $v_h^2= {2k\ep\lh}+O(\ep^2)$ and 
$v_v^2= {2k\ep\lv}+O(\ep^2)$. It then follows that $a\approx {3k\ep
\over 2\lh}+O(\ep^2 )$ and $b\approx -{\ep\over 4}+O(\ep^2)$.
Neglecting terms of $O(a\ep^2)$ and higher the modulus
potential  under this condition takes the following form

$$V(\rc )\approx {3k^2\ep^2\over 4\lh}+k\ep[a \tphi^{4+2\ep}-v_v^2
\tphi^4].\eqno(10)$$

The above expression for $V(\rc )$ exhibits a minimum at $\tphi^{2\ep}
\approx {v_v^2\over a}$ or $2\ep k\rc\pi\approx 
\ln{3 v_h^2\over 4 v_v^2}$. For $k\rc\approx$ 12 and $\ep\approx$.01
we need to adjust $v_v$ and $v_h$ so that ${v_h^2\over v_v^2}\approx$
2.8, a condition that does not involve a large hierarchy between $v_v$
and $v_h$ and hence no fine tuning. On the other hand for $\ep \approx$
.1  the minimum of the potential would occur naturally at $k\rc\approx$1.
We would like to note that the natural order of magnitude value of
  $\lv$ and $\lh$ is
$k^{-2}$. Under this condition ${v_v^2\over k^3}\approx 
O({v_h^2\over k^3})\approx
O(\ep )\ll 1$ and  the back reaction of the bulk 
scalar field on the background metric can be neglected [2, 3].
The mass of the  stabilized radion can also be derived from (10).
It can be shown that
$$m^2_{\phi}={1\over f^2}{\del^2V\over \del\tphi^2}\approx 16 \ep^3
{k^2\over f^2\lv} e^{-2k\rc\pi}.\eqno(11)$$.
If $\lv\approx k^{-2}$ then $m_{\phi}\approx\sqrt{2\over 3}\ep^{3\over 2}
ke^{-k\rc\pi}\approx .8 $ GeV.

If we let $\lh\rightarrow \infty$ keeping all other parameters fixed then 
$a\rightarrow v_h^2$. The condition for getting a mimimum at 
$k\rc \approx 12$ for $\ep \approx $ .01 then becomes ${v_h^2\over v_v^2}
\approx $ 2.1. The requirement for minimum at $k\rc\approx 12$ for
$\ep \approx.01$ therefore does not change significantly as 
$\lh\rightarrow\infty$ starting from some finite value.

We shall now determine the form of the modulus potential for $k\rc\ll $1.
Although  a minimum at $k\rc\ll 1$ may not be useful from the point
of view of generating the weak scale-Planck scale hierarchy it
would be useful for comparing the condition for minimum at
small $k\rc$ with that for large $k\rc$ particularly with regard to
their compatibility.
In this case one can expand the exponentials as a power series
in $k\rc$ and keep
only the leading terms.  We find that the constants A and B to leading 
order in $k\rc$ are given by
$A\approx -{\ep\over 4}\sqrt {\ep k\over \lh}(1-{3\over 2}x)$
and $B\approx \sqrt {\ep k\over \lh}(1+ {x\over 2})$. Here $x=\nu k
\rc \pi\ll $ 1.
The modulus potential to leading order is given by
$$V(\rc ) \approx {k^2\ep^2\over \lh}(2x^2+3-4x)-4{k^2\ep^2\over \lv}
(1-2x).\eqno(11)$$
The above expression for 
 modulus potential exhibits a local minimum for $k\rc \ll 1$
provided $\lv >2\lh$ and the minimum occurs at $x=1-2{\lh\over \lv}$.
In order to prevent any minimum from occuring at small $k\rc$ we could
choose $\lv$ and $\lh$ so that $\lv <2\lh$. This choice will still
allow a minimum at $k\rc\approx $12 without any fine tuning. 

In conclusion in this report we have studied the GW mechanism of radion 
stabilization for quadratic potential on the visible brane and quartic 
potential on the hidden brane. We have determined the form of the modulus
potential both for small and large $k\rc$.
 By choosing $v_v^2\approx {2k\ep\over \lv}+
O(\ep^2 )$ and  $v_v^2\approx {2k\ep\over \lv}+O(\ep^2 )$ we then showed 
that the minimum of the potential can be arranged to occur around
$k\rc \approx $12 without fine tuning the parameters of the model.
We have also shown that by adjusting $\lv <2\lh$ it is possible 
to exclude the possibility of any
minimum  from occuring at small $k\rc$.
Our simplified model allows us to determine the modulus potential
for finite values of $\lv$ and $\lh$ in contrast to their infinite values
assumed in the original GW model. Since our model is quite close to the
original GW model
our work
 provides some indication that it might be possible to obtain a minimum
at $k\rc \approx $ 12 for finite $\lv$ and $\lh$ even with
quartic potentials on both branes. However as mentioned earlier
 an analytic solution then 
becomes quite difficult.

Note added: While this work was in progress Ref. [4] appeared which
discusses GW mechanism with quadratic potential on both branes. This set
up is even simpler than ours and a bit more distant from the GW model.

\centerline{\bf References}

\item{1.} L. Randall and R. Sundrum, Phys. Rev. Lett. 83, 3370 (1999);
Phys. Rev. Lett. 83, 4690 (1999).

\item{2.} W. D. Goldberger and M. B. Wise, Phys. Rev. Lett. 83,
4922, (1999); Phys. Rev. D. 60, 107505 (1999).

\item{3.} O. De Wolfe, D. Z. Freedman, S. Gubser and A. Karch,
hep-th/9909134.

\item{4.} J. M. Cline and H. Firzoujahi, hep-ph/0005235.

\end